\documentclass[11pt,a4paper,onecolumn,reqno]{amsart}
\usepackage[a4paper, margin=2.3cm, bmargin=3cm]{geometry}

\usepackage{graphicx,stfloats} \usepackage{color}
\usepackage{amsmath}
\usepackage{amsaddr}
\usepackage{bm, soul}
\usepackage{mathtools}
\usepackage[colorinlistoftodos,prependcaption,textsize=tiny]{todonotes}
\usepackage{braket}
\usepackage{hyperref}
\usepackage[square,comma,sort&compress, numbers]{natbib}

\newcommand{\mrm}{\ensuremath{\textrm}}
\newcommand{\dd}{\mrm{d}}

\newcommand{\Sc}{\mrm{Sc}}

\newcommand{\Le}{\mrm{Le}}

\newcommand{\bx}{\bm{x}}

\usepackage{placeins}

\newcommand{\colorred}{\color{red}}

\renewcommand{\colorred}{\color{black}}
\renewcommand{\st}[1]{}

\usepackage[version=3]{mhchem} \usepackage{siunitx}
 \usepackage{multirow}

\usepackage{etoolbox}
\patchcmd{\pprintMaketitle}
  {\hrule\vskip12pt}
 {\hrule\vskip12pt\ifvoid\extrainfobox\else\unvbox\extrainfobox\par\vskip12pt\fi}
 {}{}

\newsavebox\extrainfobox

\usepackage{caption}
\title{Evaluation of Quadrature-based Moment Methods in turbulent premixed combustion}
\author[stfs]{Martin Pollack$^{*,1}$, Federica Ferraro$^1$, Johannes Janicka$^2$,Christian Hasse$^1$ }
\email{pollack@stfs.tu-darmstadt.de} 
\address[]{$^1$ Institute for Simulation of Reactive Thermo-Fluid Systems (STFS), Technische Universität Darmstadt, Otto-Berndt-Straße 2, Darmstadt 64287, Germany \\
$^2$Institute of Energy and Power Plant Technology (EKT), Technische Universität Darmstadt, Otto-Berndt-Straße 3, Darmstadt 64287, Germany}

\begin{document}
\pagestyle{plain}

\begin{figure}
    \centering
    \includegraphics[ scale=0.9]{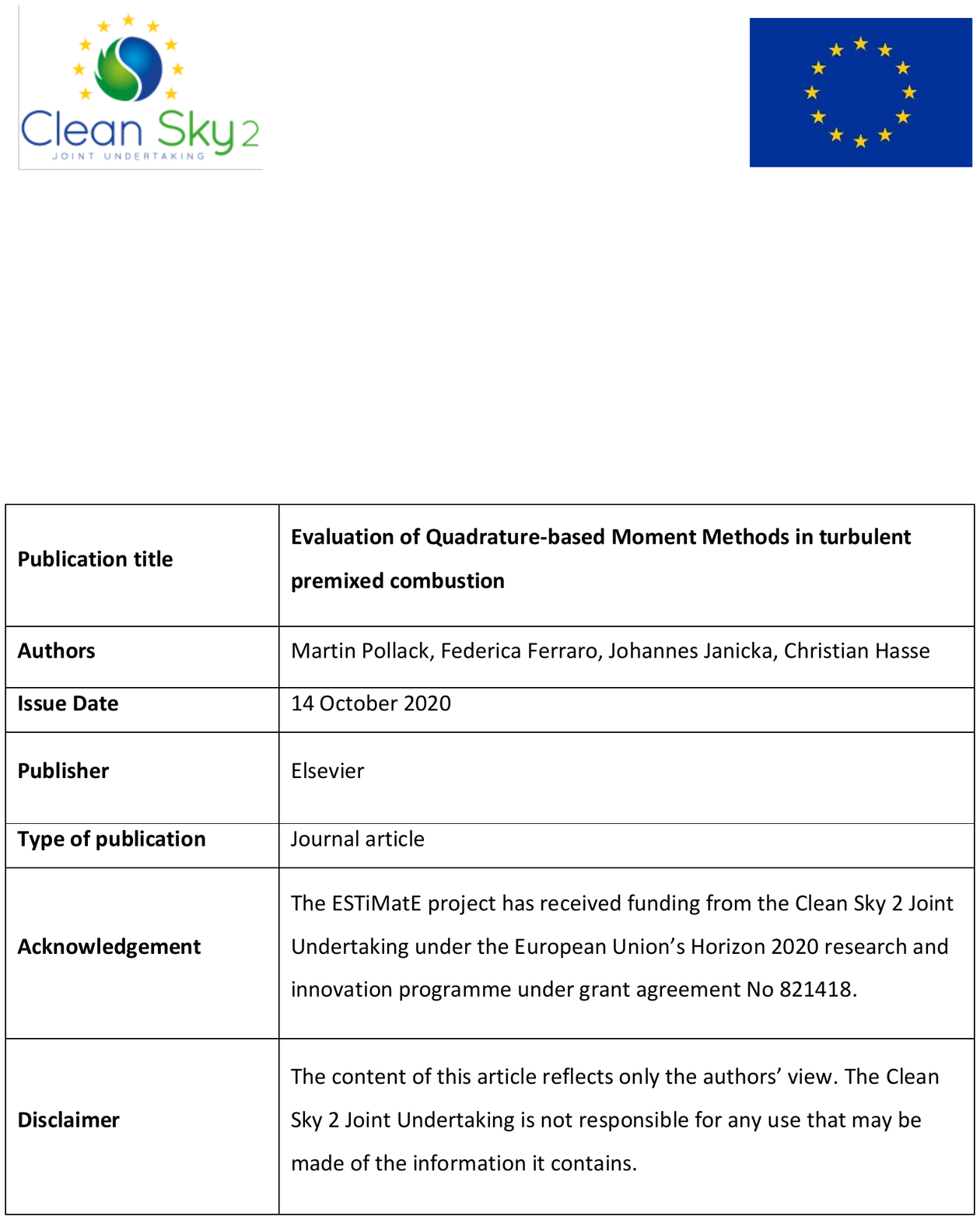}
	\caption*{}
\end{figure}
\FloatBarrier

\newpage
\maketitle

\begin{abstract} 
Transported probability density function (PDF) methods are widely used to model turbulent flames characterized by strong turbulence-chemistry interactions.
Numerical methods directly resolving the PDF are commonly used, such as the Lagrangian particle or the stochastic fields~(SF) approach.
However, especially for premixed combustion configurations, characterized by high reaction rates and thin reaction zones, a fine PDF resolution is required, both in physical and in composition space, leading to high numerical costs.
An alternative approach to solve a PDF is the method of moments, which has shown to be numerically efficient in a wide range of applications.
In this work, two Quadrature-based Moment closures are evaluated in the context of turbulent premixed combustion.    
 The Quadrature-based Moment Methods~(QMOM) and the recently developed Extended QMOM~(EQMOM) are used in combination with a tabulated chemistry approach to approximate the composition PDF.  
Both closures are first applied to an established benchmark case for PDF methods, a  plug-flow reactor with imperfect mixing, 
and compared to reference results obtained from Lagrangian particle and SF approaches.

Second, a set of  turbulent premixed methane-air flames are simulated, varying the Karlovitz number and the turbulent length scale.  
The turbulent flame speeds obtained are compared with  SF reference solutions.
Further, spatial resolution requirements for simulating these premixed flames using QMOM are investigated and compared with the requirements of SF.\\
The results demonstrate that both  QMOM and EQMOM approaches are well suited to reproduce the turbulent flame properties.
Additionally, it is shown that moment methods require lower spatial resolution compared to SF method.
\end{abstract}

\keywords{
\textbf{Keywords:} Transported PDF, Quadrature-based moment methods, Stochastic fieldss, Turbulent premixed flames }

\section{Introduction} \label{S:I}

Different approaches have been developed for modeling  turbulence-chemistry interaction. 
Among them, one very popular approach is the transported probability density function (PDF) method, first proposed in~\cite{Pope1979a}.
Here, the interaction between turbulence and chemistry is described by the evolution of a one-point joint PDF of selected flow variables \cite{Muradoglu1999}. 
In composition PDF methods, the PDF maps the likelihood of a specific state~${\colorred\bm{\phi}}$ on the composition space~${\colorred\bm{\psi}}$, usually containing the species mass fractions and  enthalpy, $\bm{\phi}=(Y_1,Y_2,\cdots,Y_{N_s}, h)^T$. The main advantage of  PDF methods is that the non-linear chemical source term appears in closed form.
The PDF~$P(\bm{x},t;\bm{\psi})$ is characterized by high dimensionality and a finite-volume discretization is impractical due to excessive numerical costs.

Different Monte-Carlo~(MC) approaches have been established for solving the PDF transport equation.
The first was the Lagrangian approach~\cite{Pope1981}, where the PDF is modeled by the evolution of a large set of stochastic particles.
An alternative PDF representation based on Eulerian MC stochastic fields (SF) was derived in~\cite{Valino1998}. In the SF method, a set of fields evolves to approximate the transported PDF.
Its Eulerian nature provides fields which are continuous
and whose mean properties are smooth and differentiable in space~\cite{Valino1998}.
This method, relatively new compared to the particle PDF approach,  
 has been applied to premixed~\cite{Avdic2016}, partially premixed~\cite{Jones2009} and non-premixed systems~\cite{Jones2010}.

The high computational effort, especially because of the chemical reactions, has limited the application of transported PDF methods to reduced mechanisms or less complex fuels. 
In~\cite{Avdic2016} the Eulerian SF method was combined with the tabulated chemistry approach, yielding a substantial reduction in the PDF dimensionality and the computational costs.

An alternative and very efficient mathematical approach to solve  PDF-based systems is the method of moments (MOM). 
Similarly to SF method, MOM  can be solved  using conventional Eulerian discretization techniques.
Here, however, only a set of integral PDF properties, i.e.\ its moments~$m_k(t,\bx)$, are solved.
Moment methods have been applied successfully to a wide range of problems, such as nano-particles and aerosols~\cite{McGraw1997}, sprays~\cite{Pollack2016}, {\colorred and combustion-related problems such as} soot formation~\cite{Salenbauch2019,wick_2016}. 
 
 Although the general idea of applying Quadrature-based Moment Methods~(QbMMs) for transported PDF systems was already proposed in~\cite{Fox2003}, QbMMs still play a minor role. Only few publications, described in the following, are available in the literature. They already indicate the great potential in combustion PDF modeling.
In the context of  QbMM formulations, different closures were derived, starting with the general univariate QMOM~\cite{McGraw1997} and later the Direct Quadrature Method of Moments~(DQMOM)~\cite{Marchisio2005}, which is considered  the first multivariate approach.
DQMOM was applied to solve the moment transport equations for the composition PDF in~\cite{Raman2006, Tang2007, Akroyd2010,Koo2011,Jaishree2012}.
However, the direct transport of nodes and weights does not satisfy the conservation rules and the approach can also become numerically unstable~\cite{Donde2012}.
Therefore, Donde et al.~\cite{Donde2012} extended this approach to Semi-discrete QMOM~(SeQMOM), combining DQMOM with a direct transport of moments. 
After recent developments, such as Conditional QMOM~(CQMOM)~\cite{Cheng2010b} for multivariate PDFs, or Extended QMOM (EQMOM), enabling continuous PDF reconstructions, PDF-QbMMs were no longer restricted to closure by DQMOM.
In~\cite{Madadi-Kandjani2017} a local, pure-mixing problem was solved using two different micro-mixing models:  interaction by exchange with the mean~(IEM) and  Fokker-Planck. EQMOM was shown to provide an accurate PDF reconstruction.
A first proof of concept that a reacting system could be closed using CQMOM was provided in~\cite{Madadi-Kandjani_phd_2017}, while the underlying general idea for the closure of different systems was additionally discussed in~\cite{Fox2018}. \\
Similarly to MC transported PDF methods, which have been primarily applied to 
\colorred
\st{non-premixed turbulent flames,}
turbulent non-premixed reacting flows, 
\color{black}
most  QbMM studies have investigated non-premixed flames~\cite{Raman2006,Koo2011,Tang2007,Jaishree2012,Donde2012,Madadi-Kandjani_phd_2017}.\\
Following these recent developments, this work aims to evaluate the QbMM methods in turbulent premixed flames using a tabulated chemistry approach. 
 Two different closures, standard QMOM and EQMOM, will be employed.
 The  QbMM framework will be  applied to simulate (i) a plug-flow reactor with imperfect mixing proposed by Pope in \cite{Pope1981} with both linear and non-linear chemistry source term, and  (ii) a set of  freely-propagating turbulent flames for different Karlovitz numbers   and turbulent length scales. 
The simulation results will be compared with Lagrangian \cite{Pope1981}  and SF  method~\cite{Valino1998,Picciani2018} results taken from the literature.

\section{Moments transport equations} \label{S:T}

 For variable density flows, 
the transport equation for the density-weighted  joint  composition PDF  {$\widetilde{P}(\bm{x},t; \bm{\psi})$ can be derived under the assumption of equal diffusivity of the species \colorred
and unity Lewis number~\cite{Pope1979a} 
\st{and high Reynolds numbers}
\color{black}
\begin{equation*}
		\underbrace{\frac{\partial \left( \bar{\rho} \widetilde{P}(\bm{\psi}) \right)}{\partial t}}_{I \; - accumulation} 
		+ \underbrace{ \frac{\partial}{\partial x_i}  \left( \bar{\rho} \tilde{u}_i \widetilde{P}(\bm{\psi}) \right)}_{II \; - macromixing}
		 + \underbrace{\frac{\partial}{\partial \psi_d}  \left( \bar{\rho} \dot{\omega}_d(\bm{\psi}) \widetilde{P}(\bm{\bm{\psi}}) \right)}_{III \; - chemical\, reactions}
\end{equation*}
\color{black}
\begin{equation}
	\begin{aligned}
		&= - \underbrace{\frac{\partial}{\partial x_i}  \left[   \bar{\rho} \langle u_i'' |\bm{\phi}= \bm{\psi} \rangle \widetilde{P}(\bm{\psi}) \right]}_{IV \;- meso-mixing} + 
		\underbrace{ {\colorred \frac{\partial}{\partial x_i}\left( D \frac{\widetilde{P}(\bm{\psi})}{\partial x_i} \right)} }_{V \; - micro-mixing}\\
		&- \underbrace{  \frac{\partial^2}{\partial \psi_d \partial \psi_{d'}} \left[  \left\langle  D \frac{\partial \phi_d}{\partial x_i}\frac{\partial \phi_{d'}}{\partial x_i} \Big|\bm{\phi}=\bm{\psi} \right\rangle    \widetilde{P}(\bm{\psi})  \right] }_{V \; - micro-mixing},
	\end{aligned}
	\label{E:PDF_EqGeneral}
\end{equation}
where 
 summation over repeated indexes is assumed. 
 Here $\rho$ is the density, $u_i$ is the velocity vector component, $\dot{\omega}_d$  is the chemical source term of the reactive scalar $d$, $D$ is the laminar diffusivity and $\bm{\psi}$ is the sample space of the composition vector $\bm{\phi}$; $\tilde{\cdot}$ denotes density-weighted Favre-averaged quantities and    $\bar{\cdot}$ {\colorred\st{denotes}} time-averaged quantities.  
All the terms on the left-hand side are in closed form, while the two terms on the right-hand side need to be modeled. 
The  meso-mixing term (IV), i.e.\ the composition-conditioned turbulent transport in physical space,  is commonly modeled by the gradient-diffusion hypothesis  $\langle u_i'' |\bm{\phi}=\bm{\psi}  \rangle \widetilde{P}(\bm{\psi}) = D_T \nabla \widetilde{P}(\bm{\psi})$, with $D_T$ the turbulent diffusivity. 
The closure of the micro-mixing term~(V), representing the molecular mixing in composition space, is obtained in this work with the IEM model~\cite{Villermaux1972}, yielding  $\left\langle D \frac{\partial \phi_d}{\partial x_i}\frac{\partial \phi_{d'}}{\partial x_i} \Big|\bm{\phi}=\bm{\psi} \right\rangle = -1/2\tau_{MM}^{-1}(\psi-\tilde{\phi})$, with  the decay time scale $ \tau_{MM}$ of the scalar fluctuations.
The IEM is widely used in combustion simulations~\cite{Fox2003}, also recently as SF closure in e.g.~\cite{Avdic2016,Jones2010}. 
\colorred
However, IEM is known to have some shortcomings~\cite{Fox2003},
i.e.\  it does not  assure local mixing in composition space  and it does not account for the dependency of the scalar dissipation rate on the chemical source term, which strongly characterizes turbulent premixed flames, as discussed in \cite{Ren2019,Zoller2013}.
Here IEM is chosen to facilitate the comparison on benchmark calculations in the following,  where the same model was applied~\cite{Valino1998,Picciani2018}. 
However, the proposed approach is not limited to this specific micro-mixing model.
\st{More advanced  closures, such as the  EMST model, which improve localness in composition space, will be applied in future work.}}\\
Further, a combination of transported PDF and manifold approaches is promising  to describe  the thermochemical state of the reacting mixture.  The main advantage is that the composition manifold can be  parametrized with a reduced number of variables, instead of the complete set including species and enthalpy, and tabulated in pre-processing~\cite{Avdic2016}. 
If a single progress variable $Y_c$ can be used to access the manifold,  this approach leads to a univariate representation of the PDF, reducing the composition space vector to $\bm{\phi}=(Y_c)$. \\
In~\cite{Valino1998} Vali{\~n}o formulated the stochastic differential equations for an ensemble of $N_{sf}$ stochastic fields $\phi^j(\bm{x}, t)$  for $j=1,\cdots,N_{sf}$, whose evolution is equivalent to the evolution of the composition PDF in Eq.~(\ref{E:PDF_EqGeneral}) 
\begin{equation}
	\begin{aligned}
	&\dd ( \bar{\rho} {\phi}^j) = - \frac{\partial \left( \bar{\rho}\tilde{u}_i {\phi}^j  \right)}{\partial x_i} \dd t + \frac{\partial}{\partial x_i}\left[ \bar{\rho}\left( D + D_T \right) \frac{\partial {\phi}^j}{\partial x_i}\right] \dd t +\\
	& \left( 2\bar{\rho}^2 D_T \right)^{\frac{1}{2}}\frac{\partial {\phi}^j}{\partial x_i} \dd W_i^j - \frac{\bar{\rho}}{2 \tau_{MM}}\left( {\phi}^j - \tilde{\phi} \right) \dd t + \bar{\rho} \dot{\omega}({\phi}^j) \dd t.
	\end{aligned}
	\label{E:PDF_EqSF}
\end{equation}
Here the IEM micro-mixing closure is applied. 
The Wiener term~$dW_i$ introduces the stochastic character into the equation, which induces the different evolution of fields. 
Consequently, this stochastic approach requires a minimum number of fields to ensure a smooth evolution of the integral properties and limit the stochastic noise. 
In contrast to  stochastic approaches, where only the ensemble describes the overall system, moment equations are  deterministic. 
The transport equation for a moment of order $k$ is obtained by applying the  definition  $\tilde{m}_{{k}}=\int_{\Re_\psi} \psi^{{k}} \widetilde{P}(\psi) \mrm{d} {\psi}$ to the composition PDF Eq.~(\ref{E:PDF_EqGeneral}):
\begin{equation}
	\begin{aligned}
		&{\frac{\partial \left( \bar{\rho} \tilde{m}_{{k}} \right)}{\partial t}} + {\frac{\partial}{\partial x_i}  \left( \bar{\rho} \tilde{u}_i \tilde{m}_{{k}} \right)} - {\frac{\partial}{\partial x_i}  \left[ \bar{\rho}\left( D + D_T  \right) \frac{\partial \tilde{m}_{{k}}}{\partial x_i} \right]}  \\
		&=  {   k \int_{\Re_{{\psi}}} \psi^{k-1} \bar{\rho} \dot{\omega}({\psi}) \widetilde{P} \mrm{d} {\psi}  } - { k \frac{\bar{\rho}}{2 \tau_{MM}}  \int_{\Re_{{\psi}}} \psi^{k-1} \left( \psi - \tilde{\phi} \right) \widetilde{P} \mrm{d} {\psi} }.
	\end{aligned}
		\label{E:SGSEqConsGeneralMultivar}
\end{equation}
Similarly to the SF formulation discussed above, the IEM model is employed. 
A micro-mixing formulation in Eq.~(\ref{E:SGSEqConsGeneralMultivar}) is also possible in terms of moments~\cite{Madadi-Kandjani2017}. The chemical source term is, however,  unclosed, 
{\colorred contrary to  the Lagrangian and SF methods,}
since it directly depends on the unknown PDF. 
Two different moment closures employed will be described in the following.\\
\vspace{-0.4cm}
\subsection{Closure with QMOM and EQMOM} \label{SS:T_Q}
The first QbMM approach to close the moment equation was provided in~\cite{McGraw1997}, and this standard QMOM is discussed first. The recent EQMOM approach, which provides more insight to the underlying PDF, is explained afterwards.

\paragraph{QMOM}
The basic idea of QMOM is to approximate integrals containing the 
\colorred
\st{NDF,} PDF,  
\color{black}
such as the chemical reaction source term in Eq.~(\ref{E:SGSEqConsGeneralMultivar}), using a set of~$N_\alpha$ weighted quadrature nodes, $\phi_\alpha$, as follows:
\begin{equation}
	\int_{\Re_{\psi}} q(\psi) \widetilde{P}(\psi) \dd \psi \approx \sum_{\alpha=1}^{N_\alpha} q(\phi_\alpha) w_\alpha,
	\label{E:Q_C}
\end{equation}
where $w_\alpha$ are the weights and $q(\psi)$ contains all terms except the PDF itself. 
Setting $q = \psi^k$,  directly yields the moments definition:
\begin{equation}
	\tilde{m}_k = \int_{\Re_{\psi}} \psi^k \widetilde{P}(\psi) \dd \psi \approx \sum_{\alpha=1}^{N_\alpha} \phi_\alpha^k w_\alpha,
	\label{E:Q_m}
\end{equation}
showing that the system of $N_{mom}$ moments (with $m_0$ equal to unity per definition, $m_1$ as the mean, etc.) is fully determined by $N_\alpha = N_{mom}/2$ pairs of nodes and weights.
In this work,  the Wheeler algorithm~\cite{Wheeler1974} is employed to compute the nodes and the weights for the PDF reconstruction.
This step is also known as moment inversion.
Based on  Eq.~(\ref{E:Q_C}), the chemical source term can now  be approximated as~$k\sum_{\alpha=1}^{N_\alpha} \phi_{\alpha}^{k-1} \bar{\rho} \dot{\omega}(\phi_\alpha) w_\alpha$.\\
Finally, the QMOM yields a discontinuous PDF representation as weighted sum of Dirac delta function
\begin{equation}
	\widetilde{P}(\psi) \approx \sum_{\alpha=1}^{N_\alpha} w_\alpha \delta(\psi - \phi_\alpha).
	\label{E:Q_P}
\end{equation}
\vspace{-0.2cm}
\paragraph{EQMOM}
Contrary, the basic idea of EQMOM~\cite{Yuan2012} is to provide a continuous PDF reconstruction   using a set of~$N_\alpha$ known Kernel Density Functions~(KDFs), e.g.\ Gamma, Beta or Gaussian distributions. 
The continuous reconstructions are essential for several applications such as spray evaporation or soot burn-off{\colorred, as shown in~\cite{wick_2016}}. 
Here,  Beta distributions are used, since they can represent a domain bounded on either side. This applies for the progress variable, which is limited  between the unburned state~$Y_{c,min}$ and the burned {\colorred \st{equilibrium}} state~$Y_{c,max}$.
To satisfy the mapping on the beta space~$\psi\,\in\,[0,1]$, the transported moments are transformed before applying them in the EQMOM algorithm~\cite{Pollack2019}:
\begin{equation}
	\colorred\tilde{m}_{k}^{[0,1]}  	\color{black}= (Y_{c,max} - Y_{c,min})^{-k} \sum_{n=0}^{k} \frac{k!}{n!(k-n)!} \tilde{m}_{n} \cdot (-Y_{c,min})^{(k-n)} \; .
	\label{E:betaTrafo}
\end{equation}
Further, it has been shown~\cite{Madadi-Kandjani2017} that this formulation is well-suited to reconstruct a univariate, pure mixing system. The PDF reconstruction using Beta EQMOM is given as~\cite{Yuan2012}:
\begin{equation}
	\widetilde{P}(\psi) \approx \sum_{\alpha=1}^{N_\alpha} w_\alpha \frac{\psi^{\lambda_\alpha -1}(1-\psi)^{\mu_\alpha-1}}{B(\lambda_\alpha,\mu_\alpha)}
	\label{E:EQ_P}
\end{equation}
with~$\lambda_\alpha=\phi_\alpha/\sigma$ and~$\mu_\alpha=(1-\phi_\alpha)/\sigma$, where $\sigma$ is a shape parameter obtained iteratively and $B$ is the beta function. 
Introducing the additional parameter $\sigma$ requires one more moment to be solved, $N_{mom}=2N_\alpha+1$. 
Since each of the~$N_\alpha$ KDFs is known, each can be represented by an arbitrary large number~$N_{\alpha,\beta}$ of secondary quadrature nodes~$\phi_{\alpha,\beta}$. 
This step allows the source term to be integrated with a fine $\psi$-space resolution~($N_\alpha \cdot N_{\alpha,\beta}$ secondary nodes compared to~$N_\alpha$ primary nodes in QMOM, Eq.~(\ref{E:Q_C})):
\begin{equation}
	\int_{\Re_{\psi}} q(\psi) \widetilde{P}(\psi) \dd \psi \approx \sum_{\alpha=1}^{N_\alpha} w_\alpha \sum_{\beta=1}^{N_{\alpha,\beta}} q(\phi_{\alpha,\beta}) w_{\alpha,\beta} \:,
	\label{E:EQ_C}
\end{equation}
\colorred
which yields the moments approximation as $\tilde{m}_k \approx \sum_{\alpha=1}^{N_\alpha} w_\alpha \sum_{\beta=1}^{N_{\alpha,\beta}} \phi_{\alpha,\beta}^k w_{\alpha,\beta}$, similarly to Eq.~(\ref{E:Q_m}) for QMOM.\\
It is worth noting that EQMOM does not assume the shape of the PDF but it reconstructs a
general PDF shape by a sum of weighted Beta-distributions.  If EQMOM uses a single Beta-KDF, the closure is the same as
the presumed Beta-PDF approach~\cite{Fox2018}.\\
\color{black}
The algorithm used here coupled with the CFD solver is~\cite{Fox2018,Madadi-Kandjani_phd_2017}:
\begin{enumerate}
	\item solve mass and momentum transport equations 
	\item update $D_T$ and $\tau_{MM}$
	\item solve only the advection and diffusion of the moments, left-hand side of Eq.~(\ref{E:SGSEqConsGeneralMultivar})
	\item invert the moments (determine the nodes and the weights) and calculate the source terms on the nodes, reconstruct the updated moments
	\item update the thermo-physical properties using the updated nodes
	\item solve the pressure equation.
\end{enumerate}
{\colorred While in this work flame extinction/ignition are not considered, it is worthwhile to note that for such cases the Strang splitting should be replaced~\cite{lu_2017}.}
Step 4 includes the solution of the system for each EQMOM node~$\phi_{\alpha,\beta}$,
\begin{equation}
	\frac{\dd \phi_{\alpha,\beta}}{\dd t} = \dot{\omega}( \phi_{\alpha,\beta} ) - \frac{1}{2 \tau_{MM}}\left( \phi_{\alpha,\beta} - \tilde{\phi} \right) ,\
	\label{E:SGSShiftNode}
\end{equation}
from~$t$ to~$t + \Delta t$. 
Analogously, replacing $\phi_{\alpha,\beta}$ with  $\phi_\alpha$  Eq.~(\ref{E:SGSShiftNode}) yields the QMOM formulation. 
As mentioned, {\colorred the  micro-mixing closure} is  possible both using nodes as in~Eq.~(\ref{E:SGSShiftNode}) or directly using moments as in \cite{Madadi-Kandjani2017}, 
before  updating the chemical source term.
As discussed in~\cite{Yuan2012}  shifting the nodes is more consistent.
Finally, replacing $q(\phi_{\alpha,\beta})$ in Eq.~(\ref{E:EQ_C}) with $\bm{\Phi}(\phi_{\alpha,\beta})$ yields the major quantities, such as composition, temperature, density  and the thermo-physical properties
$\bm{\Phi}(\phi) = (\bm{Y},T,\rho,c_p,\mu,\lambda)^T$,   in step 5. 
Each requested state is interpolated from the tabulated manifold, using {\colorred \st{the progress variable}} $\phi=Y_c$ as a parameter.

\section{Results} \label{S:R}
In the following, two test cases of increasing complexity are simulated with  QMOM and EQMOM closure and compared with reference results obtained using the Lagrangian particle {\colorred\st{model} and SF methods. }
The QbMM results are analyzed in terms of the closures applied, the number of moments solved and the grid resolution.

\subsection{Plug-flow reactor with imperfect mixing} \label{SS:R_Pope}
\begin{figure}
    \centering
    \includegraphics[ trim=0cm 0.cm 0cm 0cm, clip=true, width=67mm]{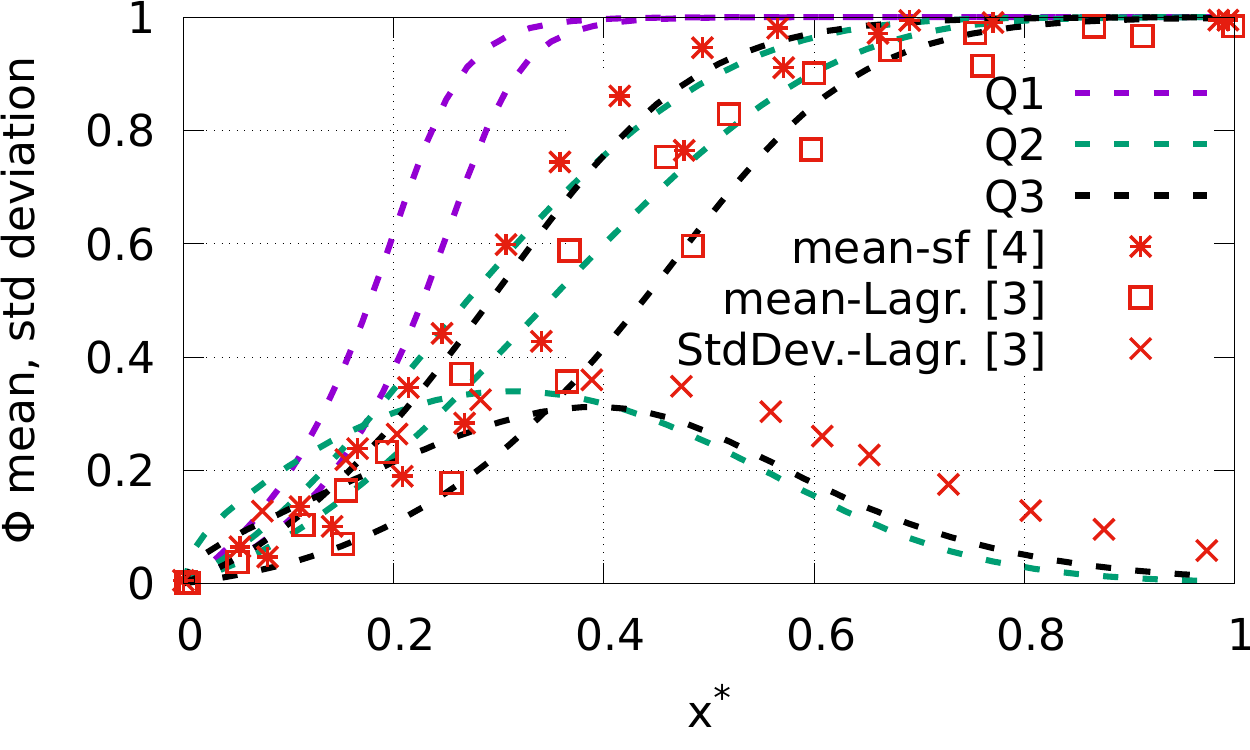}
	\caption{Mean and standard deviation profiles of the progress variable for the plug-flow reactor  at different times with Arrhenius chemical source term.
	 References  are Lagrangian~\cite{Pope1981} and SF~\cite{Valino1998} solutions.
QbMM solutions are shown for an increasing number of quadrature nodes~($Q1$ to~$Q3$). The mean curves~(square and asterisk symbols) are for $t^*=0.5$ and~$t^*=1.5$, the variance curves (crosses) for~$t^*=1$.}
    \label{F:popeNonLinQ}
\end{figure}

\begin{figure}
    \centering
    \includegraphics[ trim=0cm 0.0cm 0cm 0.0cm, clip=true, width=67mm]{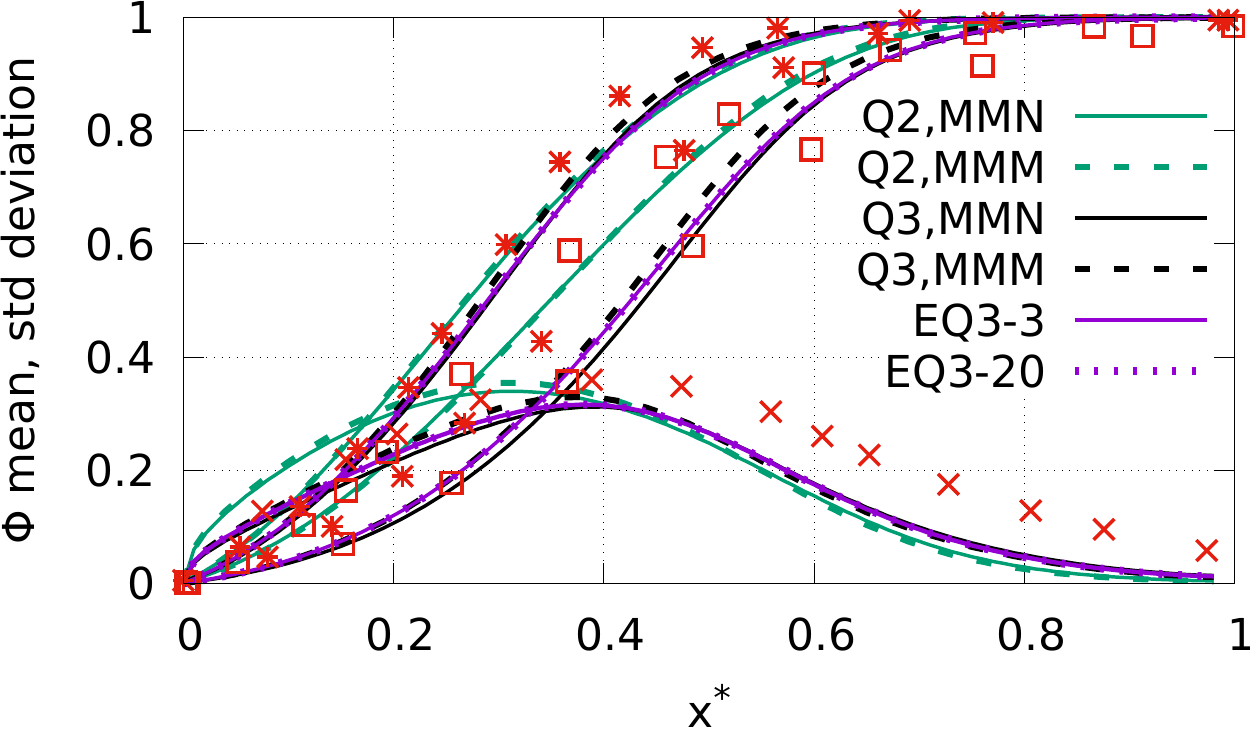}
	\caption{Mean and standard deviation profiles of the progress variable for the plug-flow reactor with Arrhenius chemical source term. Q2 and Q3 are the same as in ~Fig.~\ref{F:popeNonLinQ}.  EQMOM results are shown for $N_\alpha\!=\!3$, $N_{\alpha,\beta}\!=\!20$. Different treatment of the micro-mixing IEM closure are compared for QMOM:  closure applying directly the moments~(MMM) and closure by shifting the nodes~(MMN), as in Eq.~(\ref{E:SGSShiftNode}). Symbol legend: see Fig.~\ref{F:popeNonLinQ}. }
    \label{F:popeNonLinEQ}
\end{figure}

The test case proposed in~\cite{Pope1981} has become an established benchmark for PDF methods. It describes the evolution of a reaction progress variable $\phi$ 
in a non-dimensionalized sub-space~$t^*$, $x^*$.
Uniform non-dimensional velocity $u^*$, density $\rho^*$, diffusivity $D^*$,  chemical source term  $\dot{\omega}^*$  and micro-mixing time scale  $\tau^*_{MM}$ are considered. 
The progress variable is set to zero at the reactor inlet, $\phi=0$ at $x^*=0$ (unburned gas),  and everywhere initialized with  $\phi=1$ (burned gas). 
For the chemical reaction source term, two different  formulations are considered: a linear expression with $\dot{\omega}^* = 3(1-\phi)$ and an Arrhenius-type expression with $\dot{\omega}^*= 21830 \phi (1-\phi) \exp(-20/(1+3\phi))$.
\begin{figure}
    \centering
    \includegraphics[ trim=0cm 0cm 0cm 0cm, clip=true, width=67mm]{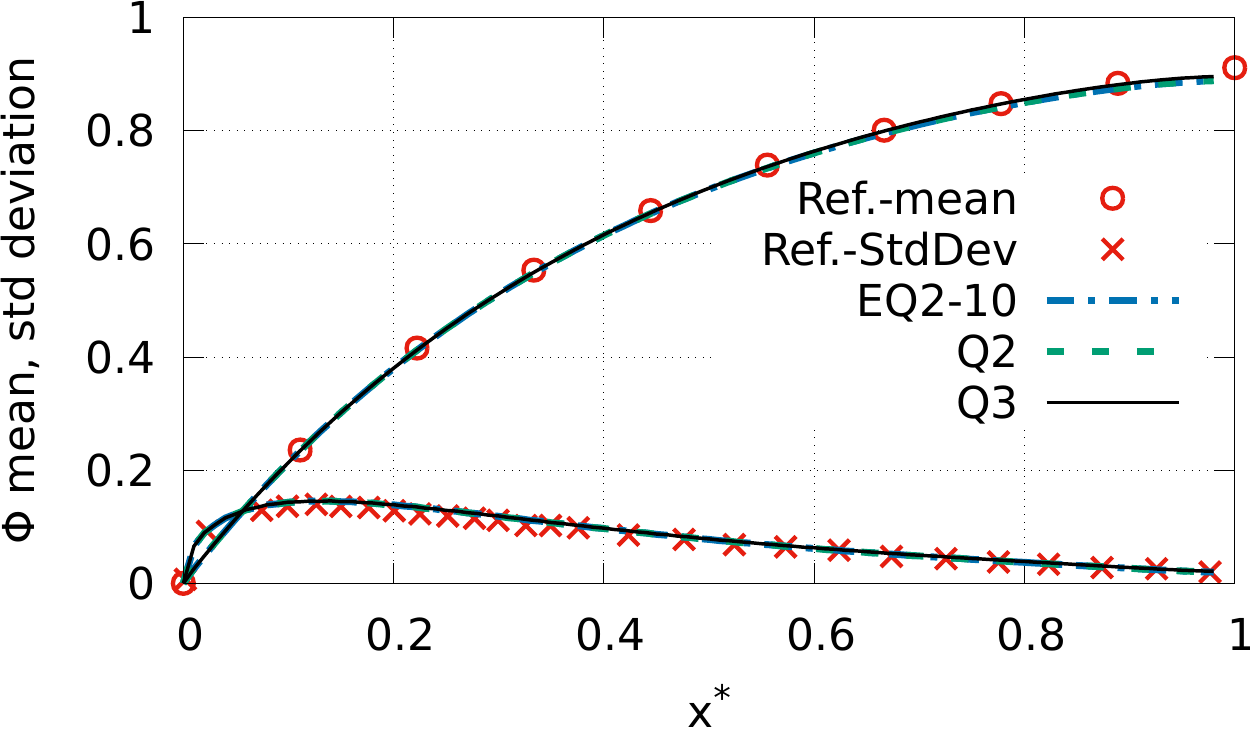}
	\caption{Mean and standard deviation profiles of the progress variable for the plug-flow reactor~\cite{Pope1981}  with linear chemical source term, compared to analytic solution (symbols) for all applied setups (lines).}
    \label{F:popeLin}
\end{figure}
Reference solutions were obtained using the Lagrangian method in~\cite{Pope1981} with 400 particles per cell and  using SF in~\cite{Valino1998} with $800$ fields.
Figure~\ref{F:popeLin} shows the  steady-state profiles for the mean and standard deviation of the progress variable over reactor length. 
For the linear source, an analytic solution~\cite{Pope1981} is  available for comparison. 
Results are shown using the standard QMOM approach with two~($Q2$) and three~($Q3$) quadrature nodes as well as with EQMOM with two KDFs  and ~$N_{\alpha,\beta}=10$~secondary nodes~($EQ2\!-\!10$).
It can be observed that, for the simple linear case, two standard QMOM quadrature nodes~($Q2$) are sufficient to provide very good results. 
According to the basic theory~\cite{Gautschi_book_2008}, the closure in Eq.~(\ref{E:Q_C}) is exact, if~$q$ is a polynomial of degree~$\le 2N_\alpha -1$, as in this case for the linear chemical source terms and micro-mixing terms.
Consequently, in the calculations ($Q3$) or ($EQ2\!-\!10$) no further significant improvement could be obtained.\\
Next, the results of the Arrhenius-type formulation are shown. In Fig.~\ref{F:popeNonLinQ}  the mean values using Lagrangian particles from Pope~\cite{Pope1981}  and SF from Vali{\~n}o~\cite{Valino1998} are plotted as references for two time instances, $t^*=0.5$ and~$t^*=1.5$~(from left to right). The standard deviation at~$t^*=1$ is taken from~\cite{Pope1981}.
The QbMM results were generated with the same numerical setup, i.e.\ the time step~($\Delta t^*=5 \cdot 10^{-4}$) and the same~$\Delta x$ resolution~($50$ {\colorred\st{nodes} cells}) as in~\cite{Valino1998} in order to ensure comparability.
Simulations considering only the mean of the distribution~($Q1$) up to higher resolved setups~($Q3$) are shown in~Fig.~\ref{F:popeNonLinQ}.
Obviously is it not possible, to reproduce this {\colorred \st{setup} configuration} with ($Q1$), since the flame propagation is drastically underestimated. The ($Q2$) setup already yields very similar results to the SF reference, but requires only 3 moments to be solved.
An increase up to~($Q3$) yields even better results than SF. They are comparable to the Lagrangian particle method results, indicating a clear convergence towards the reference data by increasing the number of moments.
The standard deviation is slightly underestimated in the downstream area for all setups. Additionally, it has been observed that  further increasing  to~4 quadrature nodes (not shown here) yields no substantial improvements in the QbMM results.
\colorred
\st{This setup demonstrates that 3~quadrature nodes are a well suited configuration and it is thus feasible to employ higher-order moments when closing this system.}
These results demonstrate that  3~quadrature nodes are a well-suited  QMOM-configuration to guarantee sufficient accuracy.  It is thus feasible to employ higher-order moments for closing this system. \\
\color{black}
Additionally, EQMOM with 3 KDFs was applied for this system 
\colorred
with 3 ($EQ3\!-\!3$) and 20 ($EQ3\!-\!20$) secondary nodes,
\color{black} 
and the results are shown in Fig.~\ref{F:popeNonLinEQ}.
\colorred 
 It can be seen, that ($EQ3\!-\!3$) and~($EQ3\!-\!20$)  curves are almost identical to each other and  with   ($Q3$) results from Fig. \ref{F:popeNonLinQ}, 
\st{It can be seen, that these curves are almost identical  with the  ($Q3$) results,} 
included here  for comparison.  Thus, EQMOM, which has the advantage of reconstructing a continuous PDF distribution of the progress variable at a slightly higher computational cost, as it is discussed below,  yields a very similar  evolution to QMOM in this configuration. 
\color{black}
This  indicates that the system is already sufficiently well resolved with standard QMOM. 
\colorred 
 \st{which is further confirmed by the two EQMOM curves for~$3$ and for~$20$ secondary quadrature nodes~($EQ3\!-\!3$ and~$EQ3\!-\!20$)  exhibiting an identical evolution.} 
 \color{black}
Instead, the numerical treatment of the micro-mixing closure has a greater effect on the results. Figure~\ref{F:popeNonLinEQ} shows this for both closures~($Q2$) and~($Q3$) using either the node-shifting representation (MMN) following Eq.~(\ref{E:SGSShiftNode}) or the direct moment-based closure~$\dot{m}_{k} = 1/2 \tau_{MM}^{-1}\left( m_{k} - m_{k-1} m_1 \right)$ (MMM), see~\cite{Madadi-Kandjani2017}.
($Q2$) is only slightly influenced by this choice, since mostly lower order moments~($m_1\cdots m_3$) are used. For ($Q3$), which considers high order moments~($m_1\cdots m_5$), the closure by shifting the nodes (MMN) shows a better agreement with the Lagrangian reference than the direct moment-based closure~(MMM).\\
{\colorred
The numerical costs of QbMM methods can be analyzed in terms of the CPU time required for the solution of all moment transport eqs.~($t^c_{momTr}$) and for the moment inversion~($t^c_{inv}$).
The latter can be seen as a computational overhead compared to other Eulerian methods~(SF or DQMOM). It, however, depends on the test case investigated.
For the local absence of a variance~(e.g.\ a fully burned- or unburned mixture), which can be directly determined by the moments~($\tilde{m}_2=\tilde{m}_1^2$), the QbMM solution is simply a single quadrature node with~$\phi_1 = m_1$, requiring a negligible computational time.
Consequently this test case, with a reaction zone spanning over the whole domain, must be considered as upper costs-limit, i.e.\ as the setup with greatest inversion effort. 
Comparing the normalized CPU time~$t^{c,*}_{inv}=t^{c}_{inv}/t^{c}_{momTr}$ for QMOM ($t^{c,*}_{inv}=1.18$ for $Q2$, $1.22$ for $Q3$, $1.30$ for $Q4$), it appears that they are comparable. For EQMOM, which is is less efficient, the ratio is about a factor~$3.8$ greater.  
Thus, the overall CPU time required by $Q3$, which provides very good results, would correspond to  the solution of~$11$ transport equations.
{\colorred While there is no predefined minimum of stochastic fields, Vali\~no~\cite{Valino1998} reported for the same case a number of~50 SFs to achieve statistical convergence.}
Consequently, it can be summarized that especially QMOM offers benefits in terms of computational efficiency, since (i)~only a few moment transport equations are solved and (ii) the inversion time is low.
}
Since the PDF-moment equations applicability has been demonstrated, the more complex turbulent methane flame is investigated next.

\subsection{Turbulent premixed methane flame} \label{SS:R_Nivarti}
\begin{figure}
    \centering
    \includegraphics[ trim=0cm 1.3cm 0cm 0.9cm, clip=true,width=65mm]{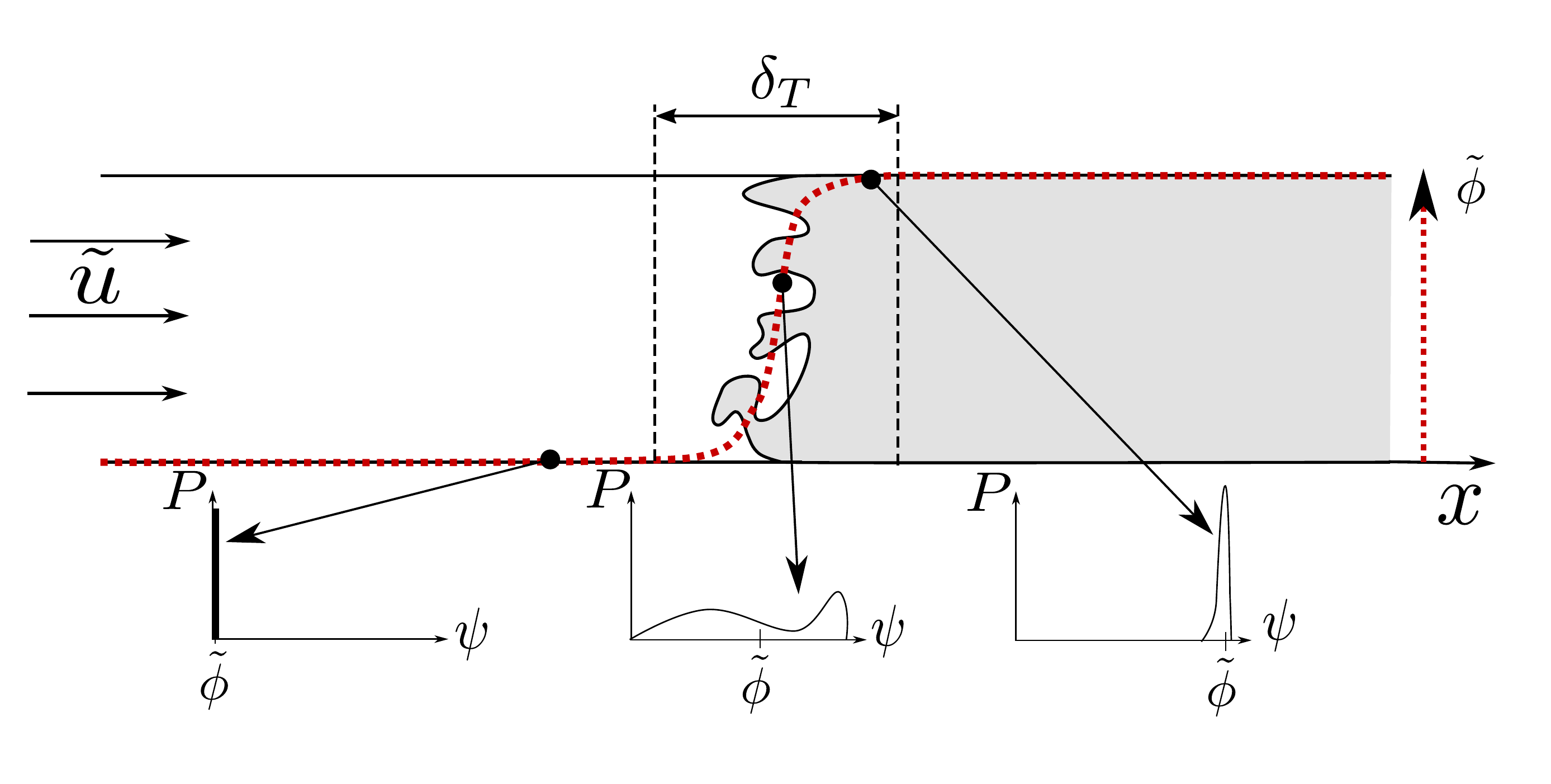}
	\caption{A sketch of a freely-propagating turbulent premixed flame, showing an instantaneous flame front, the mean progress variable profile $\tilde{\phi}$ and the turbulent flame thickness $\delta_T$. The progress variable PDF is also shown at three different locations in the flame: unburned gas (left), reaction zone (center), burned gas (right). }
    \label{F:flameScheme}
\end{figure}

\begin{figure}[!]
    \centering
    \includegraphics[width=67mm]{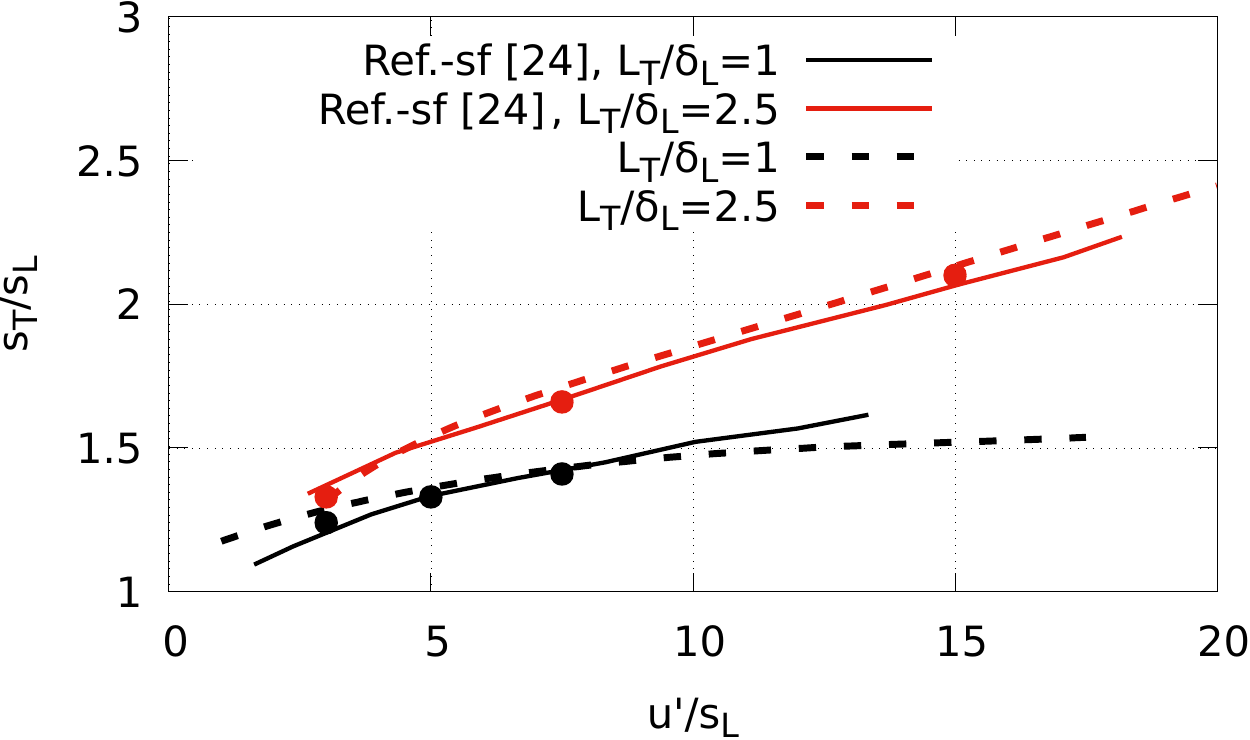}
	\caption{QMOM results  of the non-dimensionalized turbulent flame speed  for the case investigated in~\cite{Picciani2018}. The QbMM results using $Q3$ (dashed line) are compared with the SF reference using $512$~fields~(solid lines), for~$L_T/\delta_L=1$ and  $L_T/\delta_L=2.5$. Symbols indicate the cases investigated in the grid study.}
    \label{F:ch4}
\end{figure}

\begin{figure}[!]
    \centering
	\includegraphics[width=67mm]{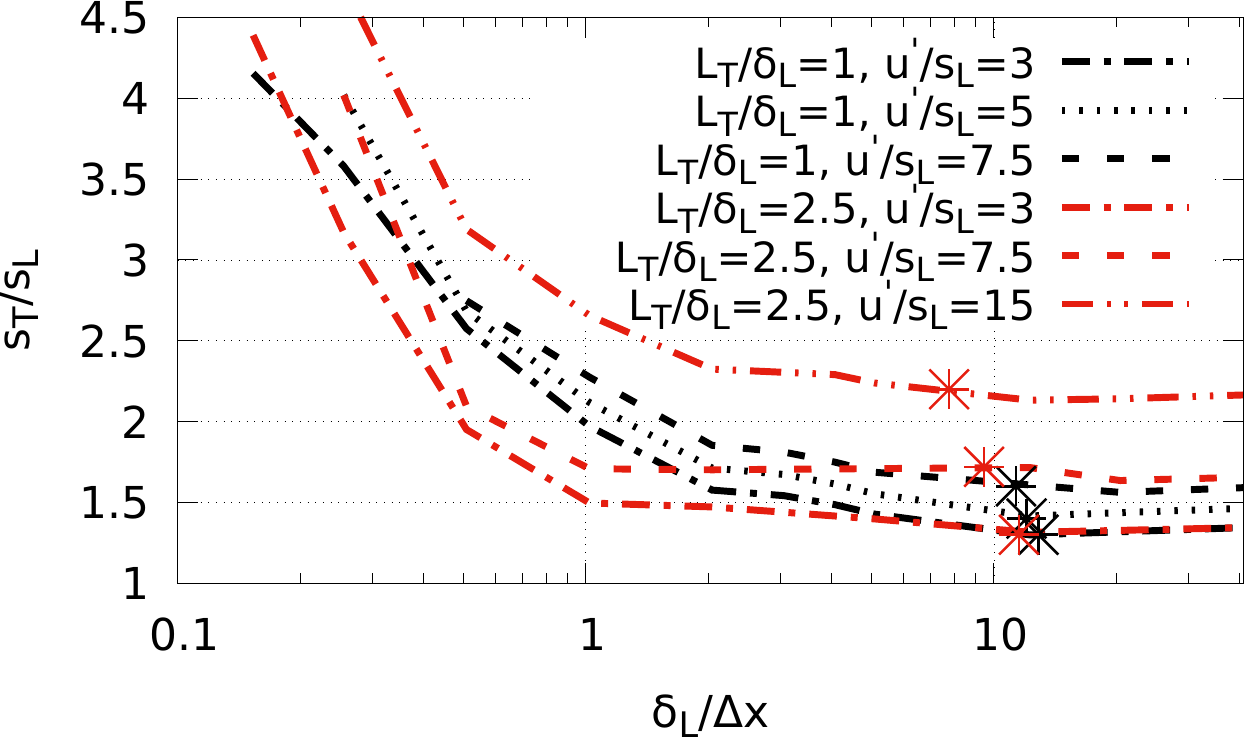}
	\caption{The turbulent flame speed dependence on the grid resolution. The left-hand side of the diagram corresponds to a coarse resolution. $\delta_L/\Delta x=1$ {\colorred\st{corresponds to} represents} a grid resolution equal the laminar flame thickness.
		For each of the cases shown, see also~Fig.~\ref{F:ch4}, the minimum grid resolution required for SF~\cite{Picciani2018} is marked by a symbol.	}
    \label{F:ch4Grid}
\end{figure}

A freely propagating turbulent  methane-air flame in stoichiometric conditions,  based on the work of Picciani et al.~\cite{Picciani2018}, is investigated in the following.
In a homogeneous isotropic turbulence, a parameter variation of characteristic turbulence properties was calculated. For different specific turbulent length scales~$L_T$, the turbulent velocity fluctuation~$u'$ was increased yielding different turbulent flame velocities and thicknesses.
This setup is based on  DNS  calculations~\cite{Nivarti2017} investigating the bending effect in turbulent flames, i.e. the reduced acceleration of the turbulent flame  at higher turbulence intensities.
Unlike the DNS, the SF-reference~\cite{Picciani2018} was solved on a 1D grid resolving the PDF with $512$ SFs. A similar setup is used in this work. A schematic representation is shown in Fig.~\ref{F:flameScheme}. 
Two different values are considered for the ratio of the  turbulent length scale and laminar  flame thickness~$L_T/\delta_L$,  namely $1$ and $2.5$. 
The ratio between turbulent fluctuations and the laminar flame speed $u'/s_L$ is varied between $1$ and $20$. 
Analogously to~\cite{Picciani2018}, the turbulent diffusivity is modeled as  $D_T = C_\mu u' L_T$ with $C_\mu=0.09$ and the micro-mixing relaxation time as~$\tau_{MM}^{-1} = (C_\phi u')/(2L_T)$, where $C_\phi\!=\!2(D_L/D_T + 1)$.
The tabulated manifold is built  solving  a freely propagating flame with the same mechanism as in~\cite{Picciani2018}, also assuming  $\Le\,=\,1$ and $\Sc\,=\,0.7$. This  leads a laminar flame speed  $s_L\!=\!0.385 \, \mrm{m/s}$ and a flame thickness $\delta_L\!=\!0.408  \,  \mrm{mm}$.
All thermo-physical properties~$\bm{\Phi}$ and~$\dot{\omega}$ are mapped on the progress variable~$Y_c=Y_{\mathrm{CO_2}}$. 
The results  for the turbulent flame speed are plotted in Fig.~\ref{F:ch4} together with the reference SF results. The turbulent flame speed and the velocity fluctuations
 are non-dimensionalized with the laminar flame speed. 
In Fig.~\ref{F:ch4} the QbMM results are based on three quadrature nodes ($Q3$). This configuration yields a good representation of the setup, which is consistent with the findings in the previous plug-flow reactor test case.
Comparing these ($Q3$) results in more detail with the SF reference shows very good agreement for both~$L_T/\delta_L=1$ and $L_T/\delta_L=2.5$, where the flame speed evolution is almost identical. \\
A point further studied for SFs in~\cite{Picciani2018} is the flame thickness and the corresponding resolution requirements. 
The QMOM simulations in Fig.~\ref{F:ch4} predict (as expected) an increase in the flame thickness $\delta_T$ over the range of  $u'/s_L$ (not shown here for brevity). 
A direct comparison with the data reported by Picciani et al.~\cite{Picciani2018}  is, however,  not possible since  only the averaged thermal thickness of the individual stochastic fields is provided in their work.
This definition based on individual fields and the flame thickness based on the progress variable mean, as provided in QMOM, can vary greatly.
However, as discussed in~\cite{Picciani2018}, the relevant flame thickness to be resolved in the context of SFs is the average thickness of all individual fields, which is lower than~$\delta_T$.
Consequently, resolving the structure of premixed flames demands  very fine numerical grids.
The grid dependency of QMOM will be analyzed and compared to SFs in the following.
Six different flames, indicated with symbols in Fig.~\ref{F:ch4},  three for $L_T/\delta_L=1$ and three for $L_T/\delta_L=2.5$ respectively, are selected and calculated for different grid resolutions.
The resulting turbulent flame speed is shown in Fig.~\ref{F:ch4Grid}, where each line represents one of these selected setups. 
The horizontal axis represents the normalized grid size, thus a value of unity corresponds to a grid resolution equal to  the laminar flame thickness.
In general, with sufficient resolution the predicted velocity is constant. 
By contrast, a non-sufficient resolution leads to an  overprediction of  the turbulent flame speed caused by large numerical diffusivity.
\colorred
It was shown in~\cite{Picciani2018} that in order to properly resolve the turbulent flames, a minimum number of 16 grid nodes are required within the average stochastic field thickness.
In Fig.~\ref{F:ch4Grid} the SF resolution is marked with a symbol for each setup and it  is generally much lower than the laminar flame thickness.
\st{The SF minimum resolutions required to properly resolve the flames given in [24] are marked with symbols for each setup. These are generally much lower than the laminar flame thickness.}
\color{black}
As expected, $L_T/\delta_L=1$ shows that a fine grid resolution is required for both methods.
Since the flame thicknesses  
are  comparable for this setup,  the resolution requirements are  also similar. 
For the flames with~$L_T/\delta_L=2.5$, however, it is seen that with QMOM $s_T/s_L$ converges with a resolution almost one order of magnitude smaller than with SFs, obviously due to the increased turbulent flame thickness.
This resolution robustness is a further advantage worth noting, also with respect to the numerical costs.
{\colorred For real applications, commonly  employing coarser grids, the applicability of this approach requires  further study.}


\section{Summary} \label{S:S}
Two QbMM closures, QMOM and (for the first time) EQMOM, were used to close the moments equation formulation of the composition PDF{\colorred, which, differently from classical TPDF methods, requires an additional closure for the chemical source term.
A tabulated manifold approach has been employed to account for chemical reactions.
\st{and combined with a tabulated manifold approach.}}
This configuration was first applied to a benchmark case, a reactive  plug-flow reactor, and compared with SF and Lagrangian particle reference results from the literature.
The influence of different numbers  of moments was tested,
with up to 4~QMOM nodes~(7 moment transport equations) and 3~EQMOM KDFs~(6 moments).
It was observed that 3 quadrature nodes or KDFs, respectively, yield very good results for all setups.
The second application, a  turbulent premixed methane-air flame, shows results which compare well with the SF{\colorred\st{reference}} results from the literature.
As with many other composition {\colorred\st{transported}} TPDF methods, a very high grid resolution is necessary for premixed flames, lower than the laminar flame thickness.
Especially for higher levels of turbulence, QbMM shows that  the flame speed can be predicted with a substantially coarser grid. 
The results and the numerical benefits, in terms of low  number of moment transport equations to be solved and the spatial resolution requirements, are very promising for the future and represent a significant advance for the use of QbMM methods for reactive flows.
%

\section*{Acknowledgments} \label{Acknowledgments}
This research has been funded by the Deutsche Forschungsgemeinschaft (DFG, German Research Foundation) - Projektnummer 237267381 - TRR 150 and by the Clean Sky 2 Joint Undertaking under the European Union’s Horizon 2020 research and innovation programme under the ESTiMatE project, grant agreement No 821418.
We further thank Dr.\ M.\ Picciani for the fruitful discussions.
\vspace{-0.35cm}

\bibliography{library.bib}
\bibliographystyle{unsrtnat_mod}

\end{document}